\documentstyle[12pt,draft,a4,epsf]{article}

\newcommand{\be}{\begin{equation}}
\newcommand{\ee}{\end{equation}}
\newcommand{\bea}{\begin{eqnarray}}
\newcommand{\eea}{\end{eqnarray}}

\newcommand{\JS}{J^{(S)}_{i,j}}
\newcommand{\JNS}{J^{(NS)}_{i,j}}

\begin{document}

\title {On the Stability of the Mean-Field Glass Broken Phase 
under Non-Hamiltonian 
Perturbations\footnote{In memory of our friend Giovanni Paladin.}}

\author{G. Iori$^{(a)}$ and E. Marinari$^{(b)}$\\[0.5em]
  {\small (a): Dipartimento di Fisica and Infn, Universit\`a di Roma
    {\em La Sapienza}}\\
  {\small \ \  P. A. Moro 2, 00185 Roma (Italy)}\\
  {\small (b): Dipartimento di Fisica and Infn, Universit\`a di Cagliari}\\
  {\small \ \  Via Ospedale 72, 09100 Cagliari (Italy)}\\[0.3em]
  {\small \tt iorig@roma1.infn.it}\\
  {\small \tt marinari@ca.infn.it}\\[0.5em]}


\maketitle
\begin{abstract}

We study the dynamics of the SK model modified by a small 
non-hamiltonian perturbation.  We study aging, and we find that on the 
time scales investigated by our numerical simulations it survives a 
small perturbation (and is destroyed by a large one).  If we assume we 
are observing a transient behavior the scaling of correlation times 
versus the asymmetry strength is not compatible with the one expected 
for the  spherical model.  We discuss the slow power law decay of 
observable quantities to equilibrium, and we show that for small 
perturbations power like decay is preserved.  We also discuss the 
asymptotically large time region on small lattices.

\end{abstract}
\vfill
\begin{flushright}
  { \tt cond-mat/9611106 }
\end{flushright}

\newpage

\section{Introduction\protect\label{S-INTROD}}

The non hamiltonian generalization of the dynamic of the SK model is 
an interesting problem. The main question is if a complex 
dynamic (implying aging, slow power like decays and related effects) 
is stable under a small non hamiltonian perturbation.

The problem has been analyzed in detail in many papers, but definite 
conclusions are difficult to reach.  Crisanti and Sompolinsky 
\cite{CRISOA,CRISOB} have studied a simplified version of the mean 
field, the  spherical model (see later in the text), and have shown that 
for this model a small perturbation is enough do destabilize the 
glassy phase. For large asymmetry two other studies agree with this 
conclusion \cite{HEGRSO,PARISI}.

The fully asymmetric model has been studied in detail in 
\cite{CRISOA,PARISI,RISCZI,SCHREC}.  The $T=0$ case has been discussed in 
\cite{PFRISC,SCHRIE,EISOPA,NUTKRE}.  Techniques like damage spreading 
have been also used \cite{ALBECA} (by reaching conclusions about the 
possible survival of a complex behavior similar to the ones we reach 
here).  Also recently the $p$-spin model (with $p>2$) has been 
studied \cite{CKLP}.

Here we will try to distinguish about two different aspects of the 
non hamiltonian generalization of the SK model.  On the one side we 
will discuss its dynamical behavior, by focusing on aging like 
effects.  On the other side we will look at the long time {\em 
equilibrium} limit of the model, on smaller lattices. 

In section (\ref{S-THEORY}) we remind to the reader some of the 
theoretical results.  In section (\ref{S-DEFINI}) we give the 
definitions we will need in the following.  In section (\ref{S-AGING}) 
we discuss aging-like effects.  In section (\ref{S-POWERL}) we show 
slow, power law decays.  These two sections deal with dynamic 
behaviors: we use large lattices and we stay far from equilibrium.  In 
section (\ref{S-EQUILI}) we discuss long equilibrium runs, on small 
lattices.  In section (\ref{S-CONCLU}) we draw our conclusions.

\section{Theory\protect\label{S-THEORY}}

Even if, as we have said, a large amount of work has been devoted to 
the study of non hamiltonian disordered models \cite{CRISOA} -- 
\cite{CKLP}, the main results about the general case of the finite $T$ 
spherical spin model with non-symmetric couplings have been obtained 
in \cite{CRISOA}, while many of the other works deal with special 
cases like the $T=0$ dynamics or the fully asymmetric case.  Here we 
are interested to the finite $T$ dynamics of a SK model (as opposed to 
the spherical model) that is perturbed by a small, non 
hamiltonian term, and as we will discuss now not much is known about 
this case.

In order to make the situation clearer let us give to the reader at 
least a sketchy idea about the results obtained in 
\cite{CRISOA,CRISOB}, that are based on the dynamic mean field 
formalism of \cite{SOMZIP}.

The generalization of the usual spin glass dynamic is done by using 
the couplings

\be
  \label{E-JJJ}
  J_{i,j} = J^S_{i,j} + k J^{AS}_{i,j}\ ,
\ee
where $J^S_{i,j}=J^S_{j,i}$, and $J^{AS}_{i,j} = -J^{AS}_{j,i}$ (with 
a definition of the asymmetry parameter $k$ slightly different from 
the $\epsilon$ we will use in the following, see (\ref{E-EPSILON})). 
Both $J$ types have zero average over the disorder and

\be 
  \overline{J^{S^{2}}} = \overline{J^{AS^{2}}} = \frac{J^2}{N} 
  \ \frac{1}{1+k^2}\ ,
\ee
where by the overline we indicate the average over the disorder and by 
the brackets we denote the thermal average.
At first one writes the non hamiltonian generalization of the {\em 
soft spin} ($\sigma_i \in [-\infty,+\infty]$) SK mean field model, 
whose dynamics is governed by the Langevin equation

\be
  \label{E-LANGEVIN}
  \Gamma_0^{-1}
  \frac{\partial}{\partial t}
  \sigma_i(t)
  = - r_0 \sigma_i(t)
  - \frac{\delta V(\sigma_i(t))}{\delta \sigma_i(t)}
  +\sum_j J_{i,j} \sigma_j(t)
  +h_i(t) +\xi_i(t)\ ,
\ee
where $J$ contains the two contributions of (\ref{E-JJJ}). The 
potential controls fluctuations in the amplitude of the soft spins 
$\sigma_i(t)$, $h$ is a local external field, and $\xi$ is a white 
noise, with

\be
 \langle
   \xi_i(t)  \xi_j(t')
 \rangle
 = \frac{2T}{\Gamma_0} \delta(t-t')\delta_{i,j}\ .
\ee
$T$ is the equivalent of a temperature (for this model, where a 
priori one cannot expect to reach thermal equilibrium). The
autocorrelation

\be
  C(t) = \overline{\langle \sigma_i(t+t') \sigma_i(t')\rangle}\ ,
\ee
and the response function

\be
 \frac
  {\delta \overline{\langle \sigma_i(t+t')\rangle}}
  {\delta h_i(t')}
  \big|_{h=0}
  \ , t\ge 0\ ,
\ee
are the main dynamical observable quantities. The static 
susceptibility $\chi$ is defined as

\be
  \chi \equiv \int_{-\infty}^{+\infty}dt\ G(t)\ .
\ee
After taking the $N\to \infty$ limit one can use the approach of 
\cite{SOMZIP} and write the mean field equations of motion

\bea
  \nonumber
  \Gamma_0^{-1}
  \frac{\partial}{\partial t} \sigma_i(t)
  = &-& r_0 \sigma_i(t)
  - \frac{\delta V(\sigma_i(t))}{\delta \sigma_i(t)}
  + h_i(t) +\phi_i(t)\\
  &+& J^2 \frac{1-k^2}{1+k^2}
  \int_{-\infty}^{t}\ dt'\ G(t-t')\ \sigma_i(t')\ ,
\eea
where $\phi_i(t)$ is a gaussian variable with zero mean and variance

\be
  \langle
    \phi_i(t) \phi_i(t)
  \rangle
    = \frac{2T}{\Gamma_0}
    \delta(t-t') + J^2 C(t-t')\ .
\ee
The detailed treatment done in \cite{SOMZIP} cannot be repeated in the 
case $k\ne 0$, where things become too complicated.  On the contrary 
the spherical model, or $p$-spin model (here with $p=2$), can still be 
treated in a satisfactory way.  The spherical model (than can be seen 
as a mean field of the mean field, or, simply, as a different model 
from SK) can be defined trough the Langevin equation

\be
  \Gamma_0^{-1}
  \frac{\partial}{\partial t}
  \sigma_i(t)
  = - r \sigma_i(t)
  +\sum_j J_{i,j} \sigma_j(t)
  +h_i(t) +\xi_i(t)\ ,
\ee
where we have only kept the quadratic part of the potential (that had 
been already extracted in the $r$-term in (\ref{E-LANGEVIN}), and 
where now the parameter $r$ is not a free parameter but such that

\be
  \frac{1}{N} \sum_i \sigma_i(t)^2 = 1\ .
\ee
In this approach one easily checks that for the symmetric model, 
$k=0$, $\chi=\frac{1}{T}$ for $T>T_g$ (where $T_g$ is the first 
temperature where $q$ takes a non-zero expectation value). There is a 
spin glass transition at $T_g=1$, and $\chi=1$ for $T\le T_g$. 

For the asymmetric case $k \ne 0$ at finite $T$ one finds that there 
cannot be any transition (we will not discuss here about what happens 
at $T=0$, since in our numerical simulations we always investigate 
the system at finite $T$): $q=0$ for all $T>0$. The reason is that 
$G(\omega)$ is singular when $\chi^2=\frac{1-k^2}{1+k^2}$, while 
$C(\omega)$ is singular at $\chi=1$. That implies that $\chi$ stops 
at $1$ in order not to violate finiteness  of correlation functions.

At last we recall that \cite{CRISOA} finds that for small $k$ the 
correlation time $\tau$ diverges like

\be
  \tau \simeq k^{-6}\ .
\ee

The way of reasoning that we have explained two paragraphs ago would 
also suggest (as found by Hertz, Grinstein and Solla in \cite{HEGRSO}) 
that the same thing happens in the full SK mean field model. 
But in the case of the SK model this is only a qualitative argument: 
for example replica symmetry breaking could change things: for example 
Crisanti and Sompolinsky \cite{CRISOA} discuss the possible appearance 
of a hierarchical distribution of large correlation times, and of a 
slow component not only in $C(\omega)$ but also in $G(\omega)$. In 
the rest of this note we will study the case of the full fledged SK 
model, in the regime where the Parisi solution characterizes the 
symmetric model. 

Cugliandolo, Kurchan, Le Doussal and Peliti in \cite{CKLP} have found 
that the result of \cite{CRISOA} is valid also when one does not 
assume a priori time translational invariance \cite{CUKU}, for a 
$p$-spin model with $p>2$.  They are interested in the behavior of the 
$p$-spin model for $p>2$ (that is non-marginal, and in some sense more 
general than the SK model).  They also find that (for $N<\infty$) in 
the $p$-spin model there are deep stable states that show a complex 
dynamical behavior even in the non-symmetric case.

At last we note that Parisi \cite{PARISI} has proposed to use spin 
glass systems with non hamiltonian perturbation as a way to build 
memories that can be confused, and has suggested that the asymmetry 
can be crucial for the process of learning.

\section{Definitions\protect\label{S-DEFINI}}

We will give here all the definitions that are relevant for the model 
discussed in this paper.  We consider an infinite range model, based on 
spin variables $\sigma_i = \pm 1$, where $i=1,\ \ldots,\ N$ labels 
site. The couplings $J_{i,j}=\pm \frac{1}{\sqrt{N}}$ with uniform 
probability (both the symmetric couplings $\JS$ and the non-symmetric 
ones $\JNS$). The usual Sherrington Kirkpatrick version of the mean 
field spin glass is defined by a probability distribution for the 
spin variables built over a Hamiltonian

\be
  P(\{\sigma\}) 
  \simeq e^{-\beta H} 
  = e^{\beta \sum_i \sigma_i \sum_j \JS \sigma_j}
  \equiv e^{\sum_i\sigma_i F_i}\ ,
\ee
where the $\JS$ are the usual quenched, symmetric random variables, 
with $\JS = J^{(S)}_{j,i}$.

We run a local dynamic, usually known as {\em Heat Bath}: each spin 
$\sigma_i$ is in turn equilibrated with the field given by the other 
spins. The probability for the new spin $\sigma_i$ to be $+1$ after 
the update is

\be
  P(\sigma_i=+1) = \frac{e^{F_i}}{e^{F_i}+e^{-F_i}}\ .
\ee
We measure the expectation value (thermal and over the disorder) of 
the internal energy at time $t$

\be
  E(t) = \frac{1}{N}
  \overline{\langle \sum_i \sigma_i F_i \rangle}\ .
\ee
We always follow two copies ($\alpha$ and $\beta$) of the system in a 
given quenched realization of the couplings, and we compute the 
overlap

\be
  q(t) \equiv \frac{1}{N} 
  \langle \sum_i \sigma^{(\alpha)}_i \sigma^{(\beta)}_i \rangle\ .  
\ee
In the non hamiltonian case one updates the spins under the field

\be
  \label{E-EPSILON}
  F_i \equiv
  \beta 
  \frac{1}{\sqrt{1-2\epsilon+2\epsilon^2}}
  \sum_j
  \bigl[
    (1-\epsilon) J_{ij}^{(S)}
    +\epsilon J_{ij}^{(NS)}
  \bigr]
  \sigma_j\equiv  F^{(S)}_i  +  F^{(NS)}_i \ ,
\ee
where now $\JNS$ is drawn independently from $J^{(NS)}_{j,i}$, and we 
have called $F^{(S)}_i$ the part of $F_i$ proportional to $J^{(S)}$ 
and $F^{(NS)}_i$ the part proportional to $J^{(NS)}$. This is the 
model we will analyze in the following. $\epsilon$ and $k$ of 
\cite{CRISOA} have the same scaling behavior when they are small, and 
play the same role.

In the case of the non hamiltonian extension of the dynamics we also 
measure separately the symmetric and the non-symmetric contributions 
to the energy, i.e.

\be
  E^{(S)}(t) = \frac{1}{N}
  \overline{\langle \sum_i \sigma_i F^{(S)}_i \rangle}\ ;\ \ 
  E^{(NS)}(t) = \frac{1}{N}
  \overline{\langle \sum_i \sigma_i F^{(NS)}_i \rangle}\ .
\ee
We define the time dependent correlation function

\be
  c(t_w,t_w+t) \equiv \frac{1}{N}\sum_{i=1}^N 
  \overline{\langle \sigma_i(t_w)  \sigma_i(t_w+t) \rangle}\ .
\ee 
We measure  the correlation functions $c(t_w,t_w+t)$ for all 
$t_w$ and $t$ of the form $2^n$, with $n$ integer. All the data 
analysis we will describe in the following will be based on the 
knowledge of the  $c(t_w,t_w+t)$ at these times.

\begin{figure}
  \epsfxsize=398pt\epsffile{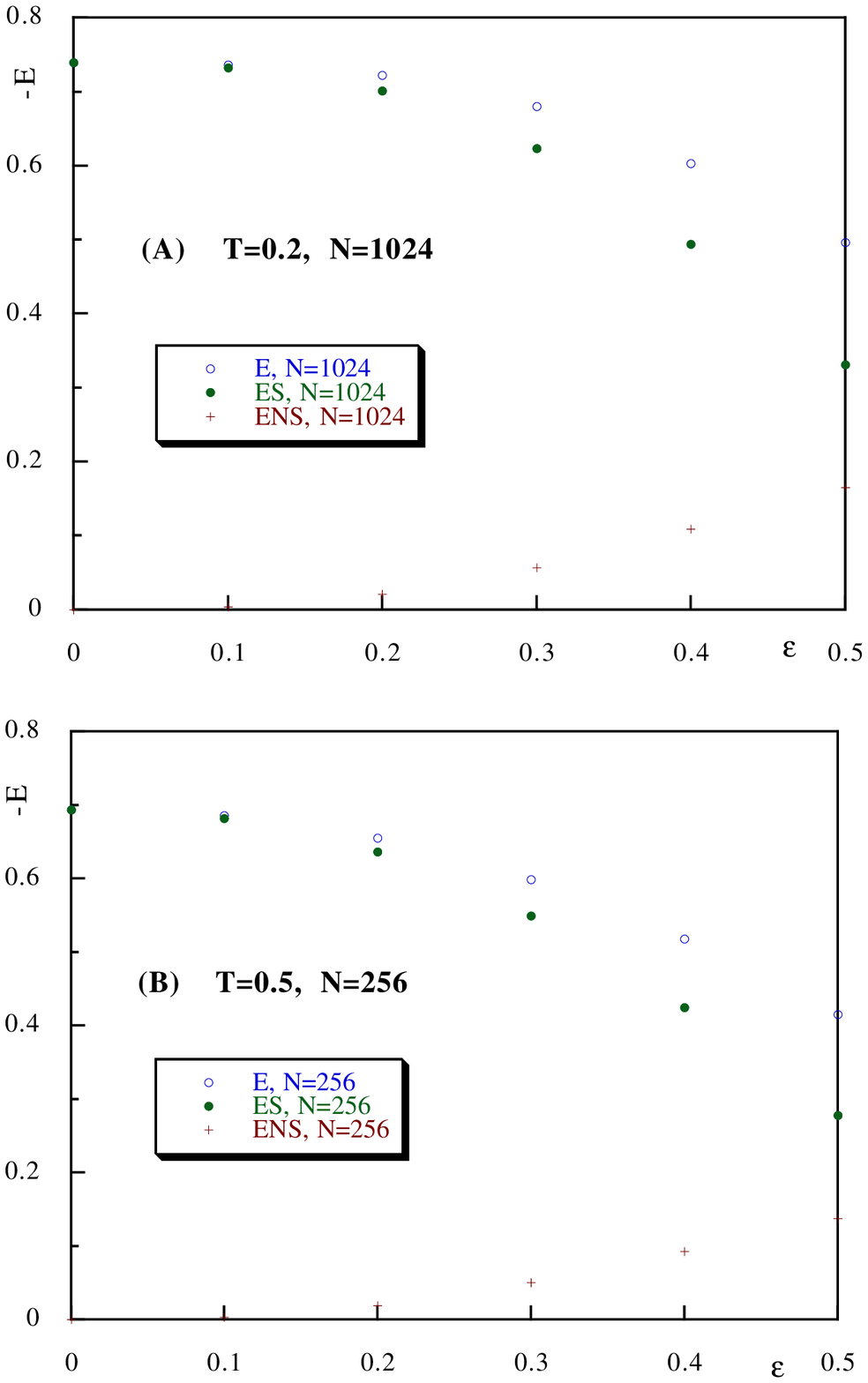}
  \caption[1]{Expectation values of the total energy operator, of the 
  symmetric part and of the non-symmetric part as a function of the 
  asymmetry parameter $\epsilon$.
  }
  \protect\label{F-FIGDE1}
\end{figure}

To try to quantify how large is the effect of a given $\epsilon$ value 
we plot in fig.  (\ref{F-FIGDE1}) the expectation value of the total 
energy operator (empty dots), of its symmetric part (filled dots) and 
of the non-symmetric part (plus symbols).  Error bars are smaller than 
the symbols.  There is an important difference distinguishing data at 
$T=0.2$ from the ones at $T=0.5$ (all taken from averages over the 
last two thirds of our total Monte Carlo sweeps).  Data at $T=0.5$ on 
a small lattice, $N=256$, are at equilibrium, in the sense that their 
average does not depend anymore on time.  On the contrary data at 
$T=0.2$ are for some $\epsilon$ values out of the equilibrium (see 
fits later on): but this is a small effect, not relevant for the point 
we are doing here.  At $\epsilon=0.0$ obviously the total energy 
coincides with the symmetric contribution, while the non-symmetric 
part is zero by construction.  As $\epsilon$ becomes different from 
zero the non-symmetric part acquires a non-zero expectation over the 
dynamics: at $\epsilon=0.1$ it stays very small, while at 
$\epsilon=0.5$ it is of the same order of magnitude than the symmetric 
contribution.  The $T=0.2$ and $T=0.5$ cases look from fig.  
(\ref{F-FIGDE1}) very similar.
  
The computer runs discussed in this paper have taken a few months of 
computers of the class of a Pentium $133$ or of a $3000/300$ Digital 
$\alpha$ Unix workstation.

\section{Aging\protect\label{S-AGING}}

At first we will try to understand the dynamical behavior of such 
systems, by studying aging phenomena \cite{RIEGER}.  We will start by 
looking, as a reference point, to the usual hamiltonian SK model.

We will look as usual at the time dependent correlation function 
$c(t_w,t_w+t)$.  Our {\em dynamical} runs (where we do not try to 
reach thermal equilibrium) have been done at $T=0.2$ (for the SK 
model, $\epsilon=0$, $T_{c}=1$).  We have studied systems with 
different number of sites, up to $N=1024$ (the case we will 
discuss in the following).  For $N=1024$ we have $20$ samples for 
each different $\epsilon$ values.

\begin{figure}
  \epsfxsize=400pt\epsffile{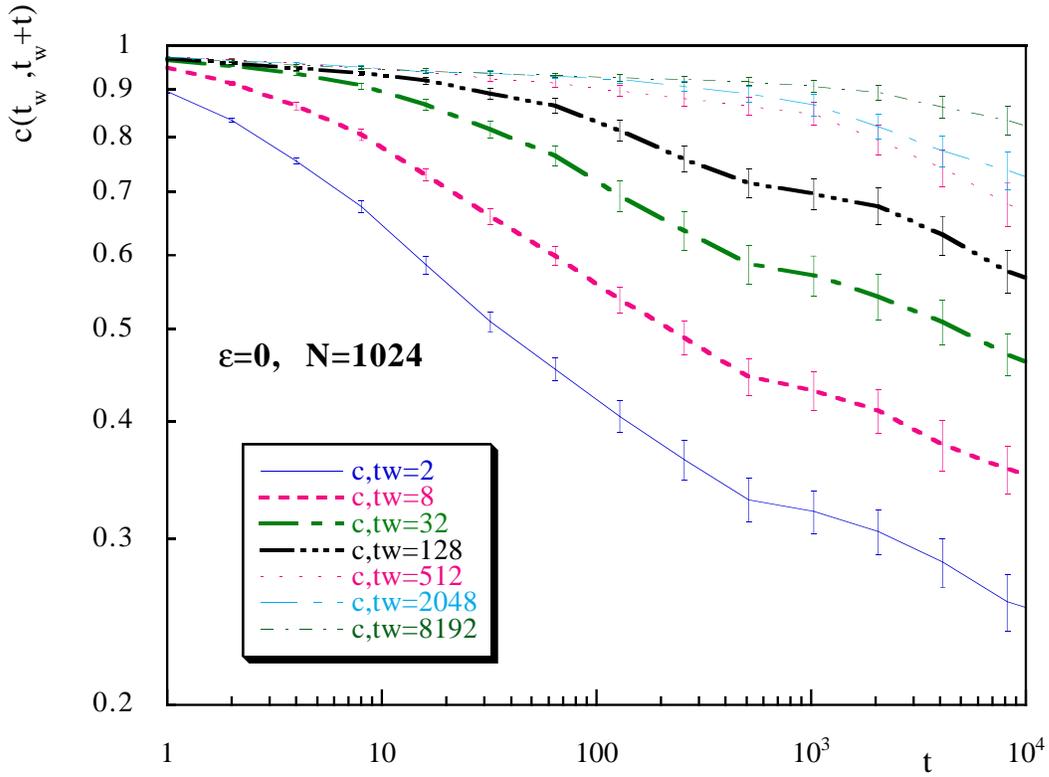}
  \caption[1]{
    The spin-spin time dependent correlation functions $c(t_w,t_w+t)$ 
    as a function of $t$, for different values of $t_w$ (lower curves 
    for smaller $t_w$ values). 
    $20$ samples, $N=1024$, $\epsilon=0$ (i.e. 
    the hamiltonian case of the usual SK model). Error bars are from 
    sample to sample fluctuations. Log-log scale.}
  \protect\label{F-FIG1}
\end{figure}

In fig.  \ref{F-FIG1} we plot $c(t_w,t_w+t)$ as a function of $t$, 
for different values of $t_w$ (lower curves depict smaller $t_w$ 
values).  Here $N=1024$, $\epsilon=0$ (i.e.  the hamiltonian case 
of the usual SK model).  Error bars come from sample to sample 
fluctuations.  The system starts from a disordered configuration, 
and we let it evolve. The fact that the correlation function is 
not time-translational invariant is very clear: $c(t_w,t_w+t)$ is 
not a function of $t$ only. For small values of $t_w$ the system 
decays fast, and the decay rate slows down when one looks at high 
values of $t_w$.

We are here in the same situation of \cite{CUKURI} for the 
simulations of the hypercubic model: on the observed time scales 
$q^2(t)$ (as measured from two copies of the system in the same noise 
realization) stays small ($\leq .04$).  The system has not crossed 
the very high barriers that is encountering on his way.  We will 
discuss that better in the followings sections, together with the fact 
that the energy is decaying to its asymptotic value with a very good 
power law (at $\epsilon=0$ and $N=1024$ the exponent is of order 
$0.4$).

\begin{figure}
  \epsfxsize=400pt\epsffile{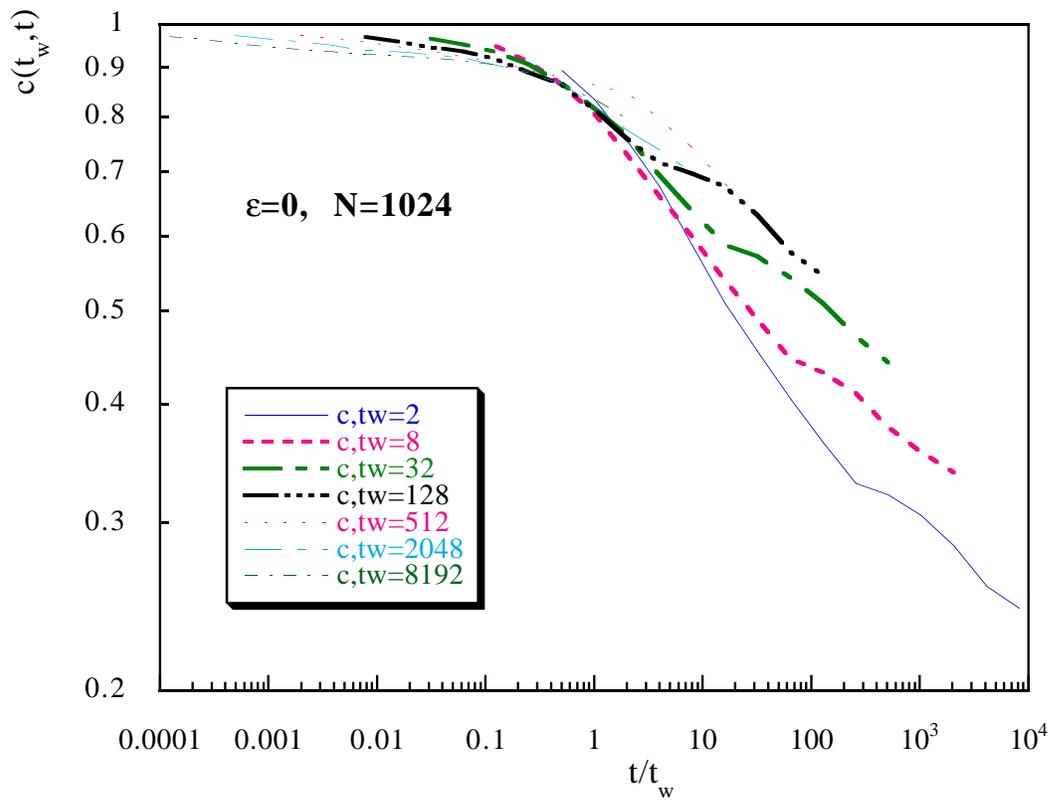}
  \caption[1]{
    As in fig. \protect\ref{F-FIG1}, but versus $\frac{t}{t_w}$.}
  \protect\label{F-FIG2}
\end{figure}

In fig.  \ref{F-FIG2} we plot the same data as a function of 
$\frac{t}{t_w}$ (again in double log scale).  This is the usual search 
for an aging-like scaling.  Here the scaling of the data is reasonable 
but obviously not perfect: scaling violations are clearly present, but 
a detailed study of this phenomenon is beyond the scope of this study.  
The plateau (to be) for $t<t_w$ is connected to the value of the same 
state overlap $q_{EA}$\footnote{We thank Juan Ruiz-Lorenzo for an 
interesting conversation about this subject.} (see \cite{LIBRO} and 
figures $3$ to $5$ of \cite{PARUTE}).  We remark that these results 
are compatible with the ones obtained by Rossetti \cite{ROSSET} on 
very large lattices (SK model up to $N=8192$) and (when looking in 
details at the graphs: the aging curves tend to separate soon after 
the crossing point for all values of $t_w$) with the ones obtained by 
Cugliandolo, Kurchan and Ritort on the alternative, hypercubic 
definition of the mean field \cite{CUKURI}.  We just repeat that 
violations of a perfect aging scaling are here quite clear.  
Definitely the usual Sherrington Kirkpatrick model does not show any 
easily explainable form of scaling.

\begin{figure}
  \epsfxsize=400pt\epsffile{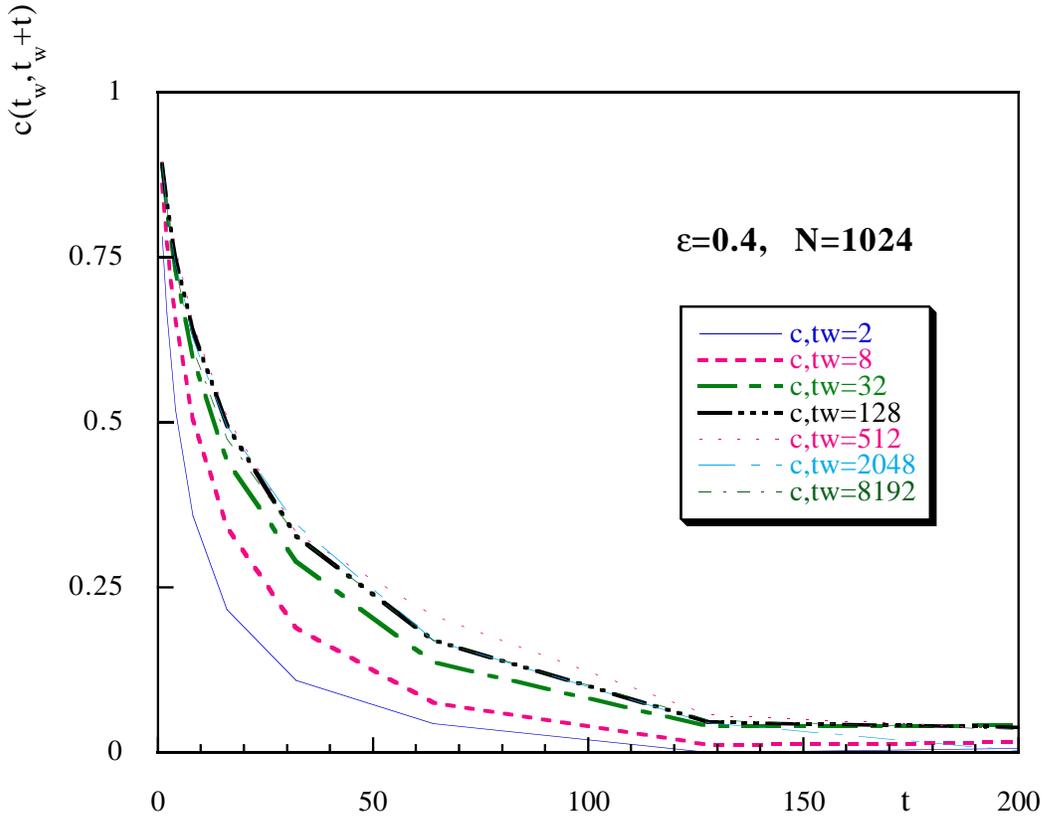}
  \caption[1]{
    As in fig. \protect\ref{F-FIG1}, but $\epsilon=0.4$, and 
    linear-linear scale.}
  \protect\label{F-FIG3}
\end{figure}

We will try now to analyze what happens in the non hamiltonian 
dynamics.  We start with the non hamiltonian dynamics by looking at a 
large perturbation: we use $\epsilon=0.4$, $20$ samples and $N=1024$.  
In fig.  \ref{F-FIG3} the correlation functions for different values 
of $t_w$ (here in simple linear-linear scale).  It is clear that 
things are now very different.  The decay at very small waiting times 
is faster than for larger $t_w$, but already for $t_w \le 32$ there is 
very little dependence of $c$ over $t_w$.  The curves from different 
$t_w$ are collapsing on a same universal decay curve.

\begin{figure}
  \epsfxsize=400pt\epsffile{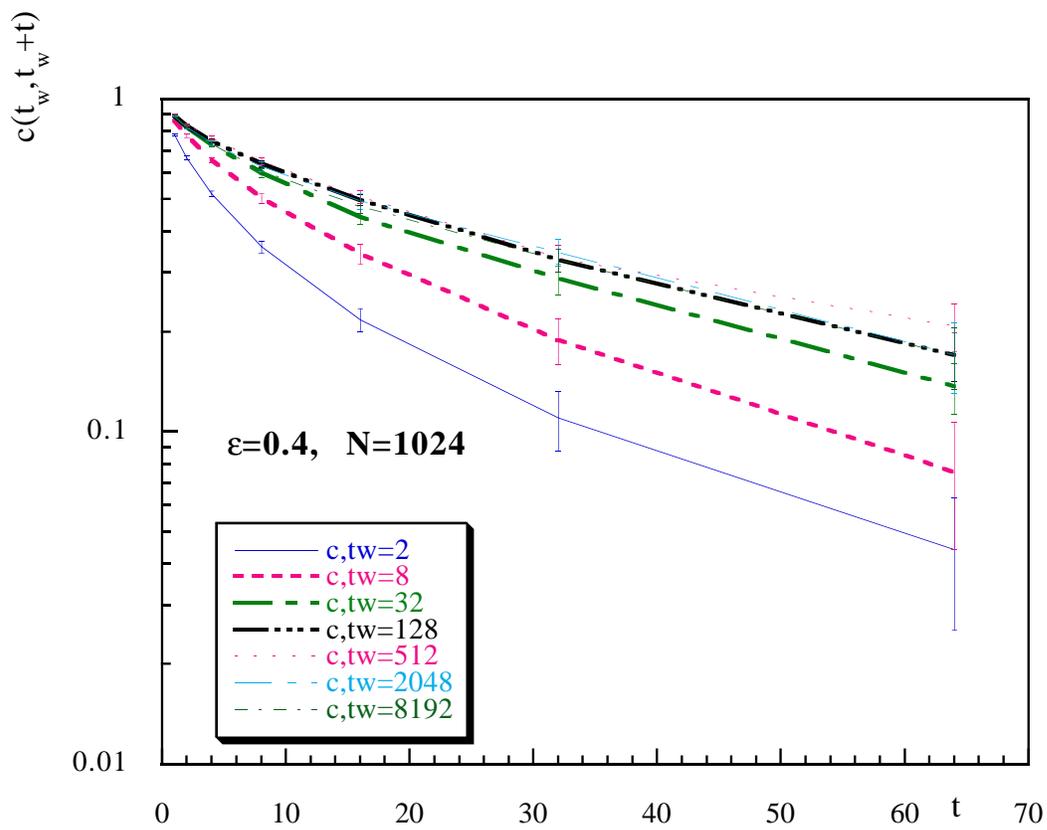}
  \caption[1]{
    As in fig. \protect\ref{F-FIG1}, but $\epsilon=0.4$, and 
    linear-log scale.}
  \protect\label{F-FIG4}
\end{figure}

Since we are maybe expecting here an exponential decay of the time 
dependent correlations (but we will see this is a far from evident 
fact) we plot in fig.  \ref{F-FIG4} the correlation functions for 
different values of $t_w$ in linear-log scale.  Here we also include 
the error from sample to sample fluctuations.  The lines would be 
asymptotically straight lines in case of an asymptotic exponential 
decay.  The careful reader can start to observe that in this case 
($\epsilon=0.4$) in all the time range where we can determine an {\em 
effective} $t$ dependent correlation time $\tau_{e}(t)$ (by looking 
for example at the local slope of the logarithm of the correlation 
function), such $\tau_e(t)$ is increasing (i.e.  the curves are 
bending up in all the region where we have been able to determine them 
with good statistical precision).

Other data for $\epsilon$ going from $.3$ to $.5$ give very similar 
hints. A first conclusion can be reached 
at this point: systems with large non hamiltonian perturbations do 
indeed have a quite different, typically non-aging, dynamical 
behavior, but one has to be careful since it looks difficult to pin 
point a real exponential decay.

\begin{figure}
  \epsfxsize=400pt\epsffile{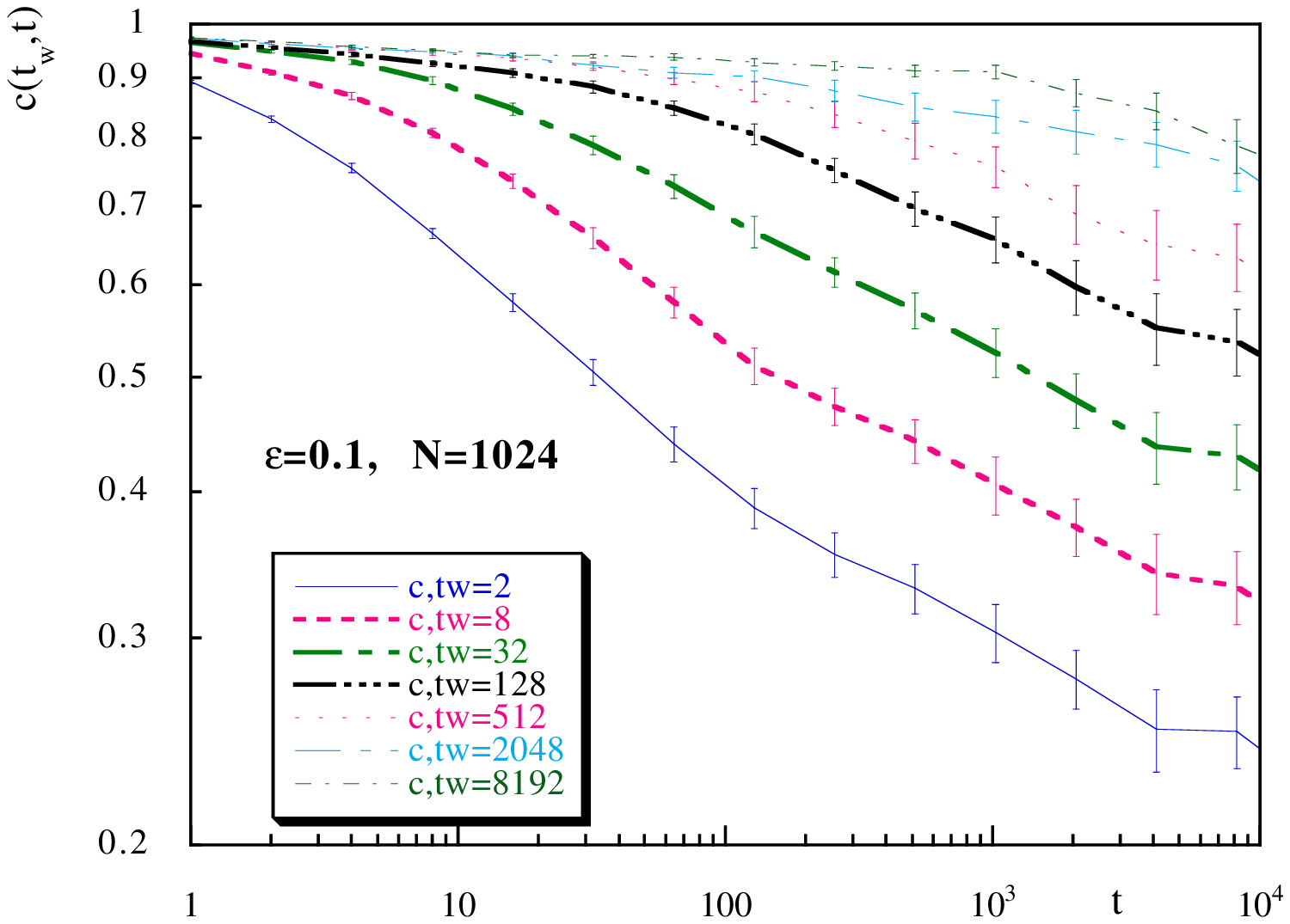}
  \caption[1]{
    As in fig. \protect\ref{F-FIG1}, but $\epsilon=0.1$.}
  \protect\label{F-FIG5}
\end{figure}

\begin{figure}
  \epsfxsize=400pt\epsffile{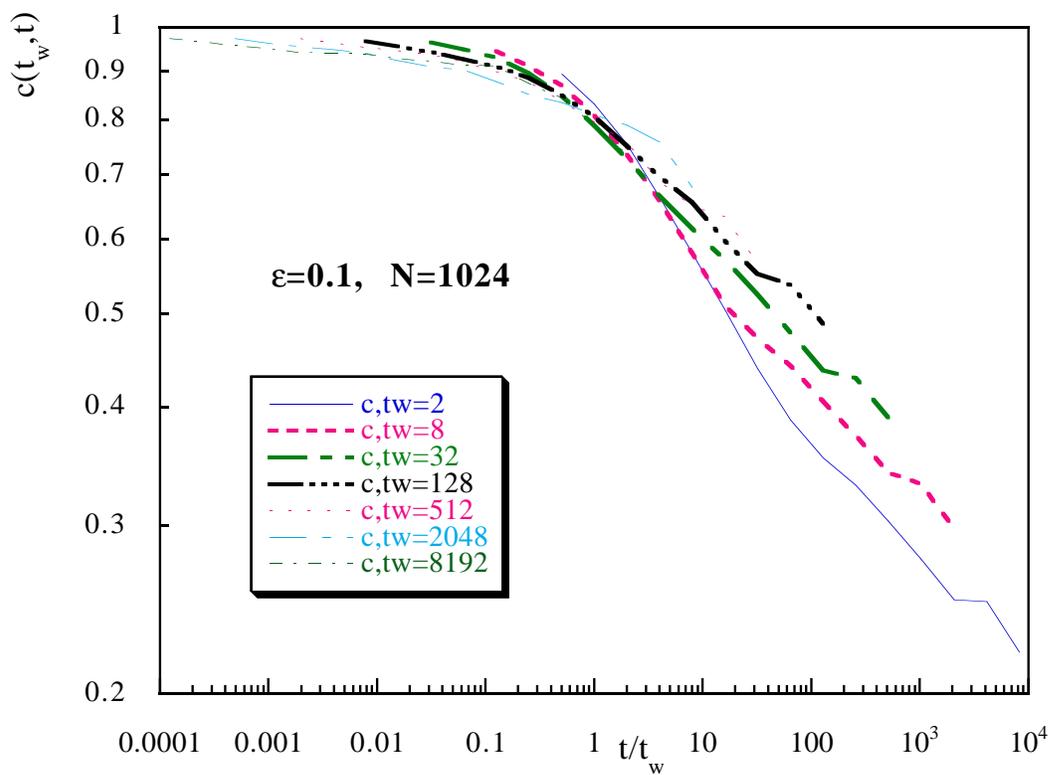}
  \caption[1]{
    As in fig. \protect\ref{F-FIG2}, but $\epsilon=0.1$.}
  \protect\label{F-FIG6}
\end{figure}

Let us now turn to a reasonably small value of the non hamiltonian 
perturbation.  We select $\epsilon=0.1$, that is not too small: the 
non-symmetric couplings are coupled with a strength $.1$ times $\beta$ 
while the symmetric couplings interact with a strength of $.9$.  These data 
turn out indeed to be dramatically similar to the ones at $\epsilon=0$.
In figures \ref{F-FIG5} and \ref{F-FIG6} the analogous of figures 
\ref{F-FIG1} and \ref{F-FIG2}: the similarity is self-explanatory. 
Later on we will try to quantify the differences and to elucidate 
their meaning. By now we observe that for large lattices ($N=1024$ 
for the infinite range model is a large lattice, since one has one 
million of couplings) and long times (we follow the system up to $t$ and 
$t_w$ of order $16384$) the model with $\epsilon=0.1$ shows an aging 
behavior as good as the one ever observed for the pure SK model.

Before discussing a more quantitative analysis of these data, a few 
comments about an intermediate case, $\epsilon=0.2$. Here the pattern 
of $c(t_w,t_w+t)$ is different from both the usual aging case 
(like the one we find for $\epsilon=0.0,0.1$) and the typical fast 
decay to an equilibrated state (see $\epsilon=0.4$, even if we will see 
that even here things are more complex).

\begin{figure}
  \epsfxsize=400pt\epsffile{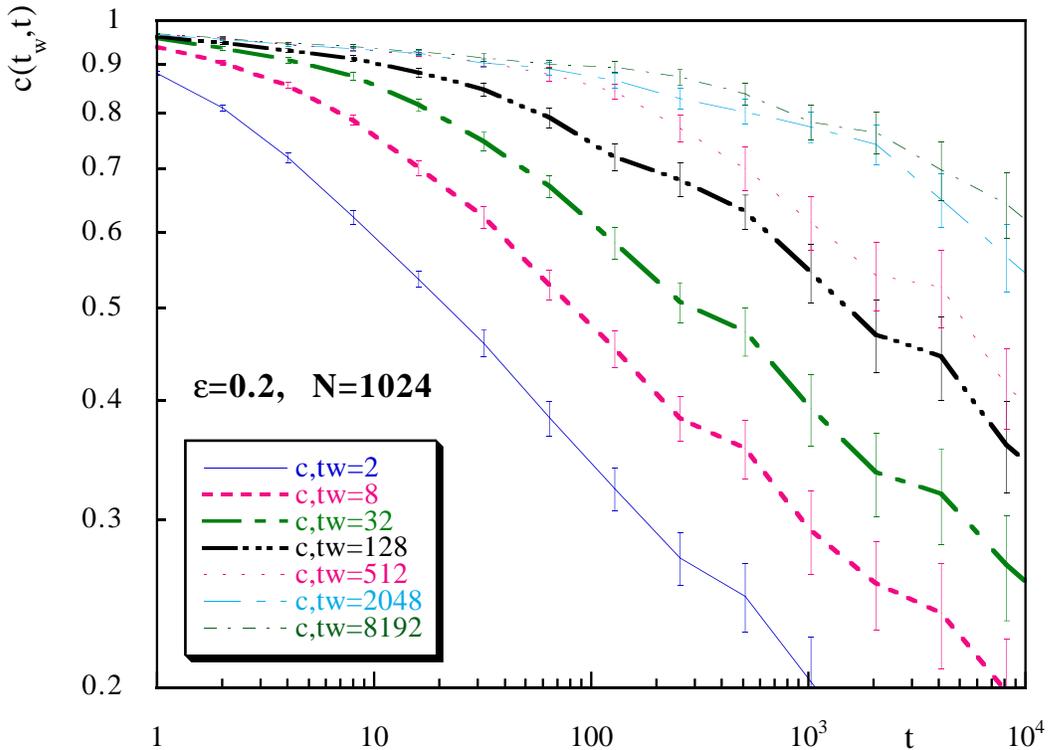}
  \caption[1]{
    As in fig. \protect\ref{F-FIG1}, but $\epsilon=0.2$.}
  \protect\label{F-FIG7}
\end{figure}

\begin{figure}
  \epsfxsize=400pt\epsffile{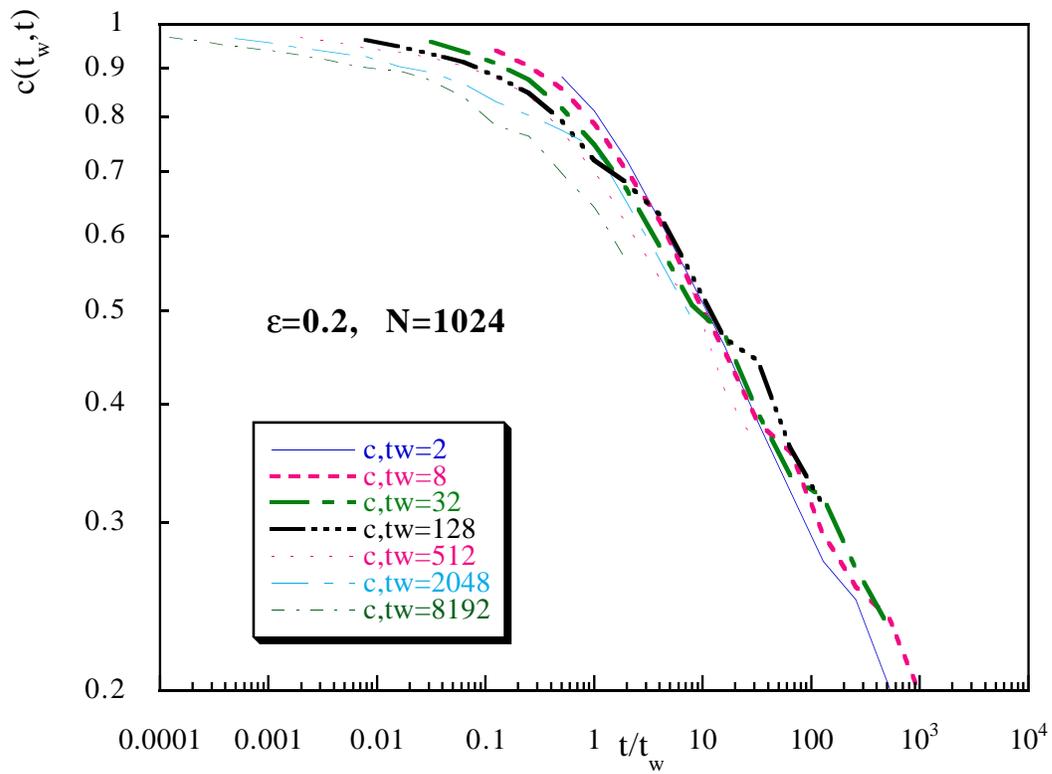}
  \caption[1]{
    As in fig. \protect\ref{F-FIG2}, but $\epsilon=0.2$.}
  \protect\label{F-FIG8}
\end{figure}

In figure \ref{F-FIG7} the correlation function as a function of $t$. 
The pattern is reminiscent of an aging pattern, even if for large 
$t_w$ the situation is not so clear. The plot of $c$ versus 
$\frac{t}{t_w}$ in fig. \ref{F-FIG8} is more innovative: here the 
typical crossing one expects at $\frac{t}{t_w}\simeq 1$is moved to 
larger values of  $\frac{t}{t_w}$, which increase with $t_w$. The 
scaling is different from the one of the hamiltonian case.

It is interesting to note, as a basis of a discussion we will present 
in one of the next paragraphs, that this is not only an effect we 
detect at high $t_w$ values.  Already for very short $t_w$ ($2$ and 
$8$, for example) we find a crossing at $t$ of order $10$.  We also 
want to notice that, looking at their value, in the region $t>t_w$ 
the $\epsilon=0.2$ data seem to show a better pure $\frac{t}{t_w}$ 
scaling than the pure ones: but this is probably a non very relevant 
feature (that could be typical of a transient behavior).

\begin{figure}
  \epsfxsize=400pt\epsffile{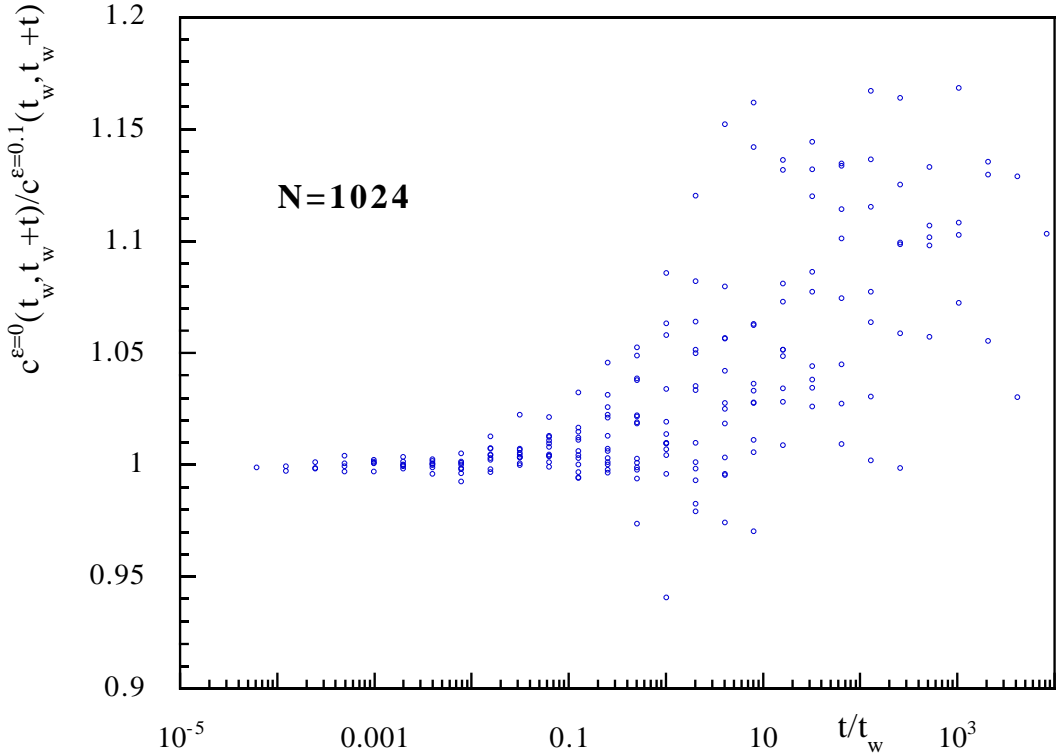}
  \caption[1]{
    The ratios of  $c(t_w,t_w+t)$ at the same $t$ and $t_w$ with $\epsilon=0$ 
    and $\epsilon=0.1$ versus $\frac{t}{t_w}$ in linear-log scale.}
  \protect\label{F-FIG9}
\end{figure}

The similarity of the correlation functions in the pure SK model 
and the ones in the model with $\epsilon=0.1$ calls for a better 
scrutiny.  The very remarkable similarity of the two sets of 
functions has to be quantified in some way.  In order to do that 
we compute the ratios of the correlation functions at the same 
$t$ and $t_w$ with $\epsilon=0$ and $\epsilon=0.1$, and we plot 
them in fig.  \ref{F-FIG9}.  The correlation functions are indeed 
very similar, but for increasing values of $\frac{t}{t_w}$ we 
start to see a small difference (typically the higher points at 
fixed $\frac{t}{t_w}$ are for higher $t$ values).  We do not plot 
the statistical errors, which would blur the plot in an extreme 
manner, but even if any individual ratio is compatible with one,
the large time growth of the ratios is clear and statistically 
significant: at large times the correlations for $\epsilon=0.1$ 
case start to decrease faster than for the pure SK case.  It is a 
very small effect, and its meaning is not unambiguous: it could 
signify that we are exiting a transient phase and aging is ending, 
or it could just be an irrelevant renormalization (a shift of the 
effective temperature of the system).  But the effect is clear, 
and we note it here.

In fig. \ref{F-FIG9} we plot the same ratio, but here $\epsilon=0$ is 
divided times   $\epsilon=0.2$. Note that the vertical scales of this 
and the former plots are very different. Here the faster decay of the 
non hamiltonian case is very clear: for $\frac{t}{t_w}$ of order 
$10^4$ the ratio is of order two. 

\begin{figure}
  \epsfxsize=400pt\epsffile{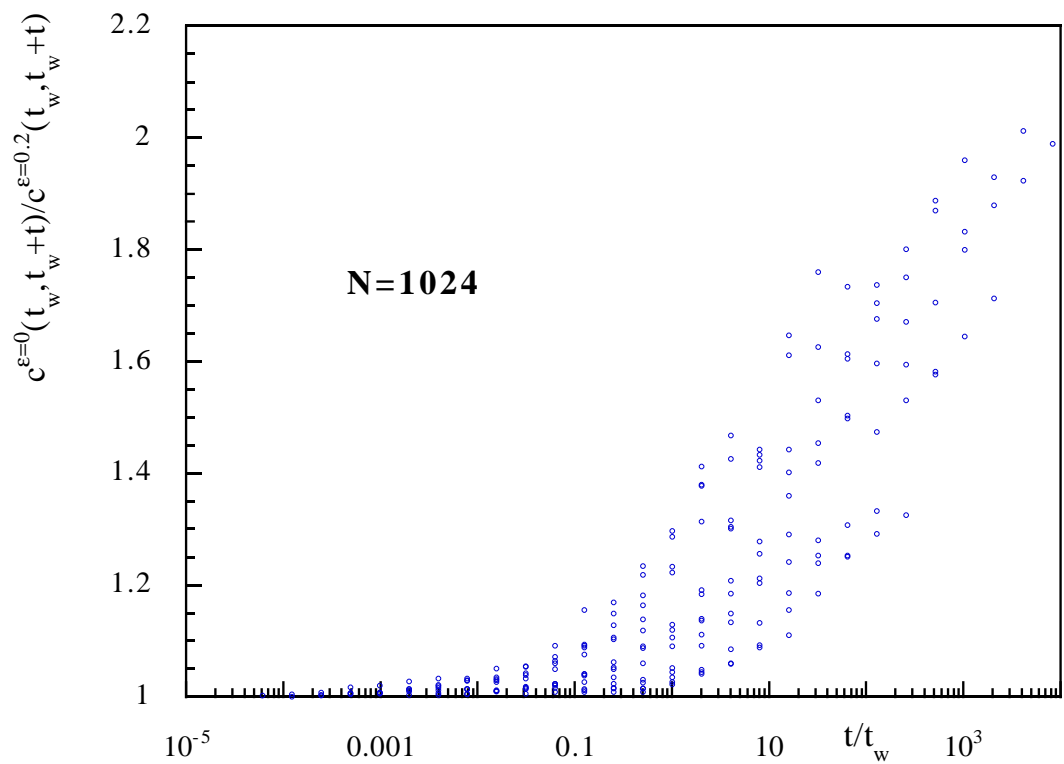}
  \caption[1]{
    As in  fig. \protect\ref{F-FIG9}, but $\epsilon=0$ 
    over $\epsilon=0.2$.}
  \protect\label{F-FIG10}
\end{figure}

Two remarks: first the departure from a ratio close to unit comes for 
the $\epsilon=0.1$ case quite late.The departure from unit is far more 
pristine at $\epsilon=.2$. For example one can notice that a value of 
$1.05$ is reached at $t/t_w\simeq 10$ in the first case, and of 
order $10^{-3}-10^{-2}$ in the second case.

If one tries in this way a very qualitative definition of an 
$\epsilon$ dependent correlation time one gets a divergence that is 
faster than the one one would get with an $\epsilon^{-6}$ scaling (see 
later).  It should also be noticed that this effect is not dominated 
by finite size effects: on different volumes we get very similar 
results.

The next natural thing to do is to check for correlation times.  The 
Crisanti-Sompolinsky \cite{CRISOA,CRISOB} result for the spherical 
model hints for a behavior

\be
  \tau \simeq \epsilon^{-6}\ ,
\ee
and one can proceed by trying to support or falsify this expectation. 
In this case one would start from the larger $\epsilon$ values, where 
the asymptotic decay looks better exposed, and try to go down to 
lower $\epsilon$ values.

Determining correlation times is typically not as easy as one expects.  
The main caveat is indeed that in a numerical simulation at her best 
one can establish upper bounds on correlation times: correlation times 
that are larger than the simulation time cannot be detected.  In the 
case of systems which exhibit (or could exhibit) an aging behavior 
things are even more complex since a priori one cannot average over 
$t_w$. Only after checking that we are in a region of time 
translational invariance we are allowed to average over different 
waiting times.

\begin{figure}
  \epsfxsize=400pt\epsffile{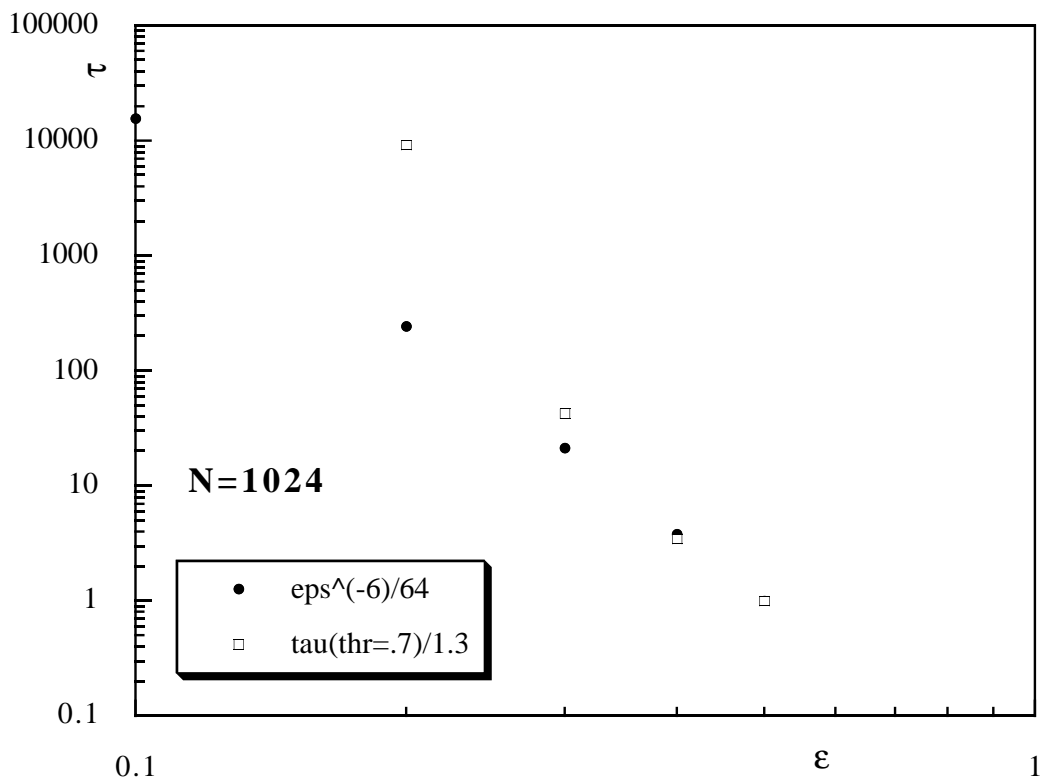}
  \caption[1]{
    As a function of $\epsilon$, in double log scale, we plot the 
    time needed from $c$ to reach the value $.7$, and 
    $\epsilon^{-6}$. 
    }
  \protect\label{F-FIG11}
\end{figure}

The first, easy approach, consists in defining the correlation time as 
the time after which the time dependent correlation function reaches, 
down from $1$, a given value, say $\tilde{c}$.  In fig.  \ref{F-FIG11} 
we plot the time needed for $c$ to reach the value $.7$, and 
$\epsilon^{-6}$ as a function of $\epsilon$ (we have also done the 
same analysis for $\tilde{c}=.5$).  It would be incorrect to average 
over $t_w$'s: only a posteriori, after checking that we are in an 
asymptotic region of a simple phase where time translational 
invariance holds, that would be justified.  We have checked indeed 
that doing that can lead to a false scaling.  In fig.  \ref{F-FIG11} 
we have only used the largest $t_w$ points available to us 
($t_w=2^{14}$).  We have normalized the data in the plot such that the 
data points coincide at the higher $\epsilon$ value ($\epsilon=0.5$).  
It is clear that already at $\epsilon=0.3$ the $\epsilon^{-6}$ scaling 
does not hold (we are using a double log scale!).  At $\epsilon=0.2$ 
the discrepancy from an $\epsilon^{-6}$ scaling is severe, and for 
$\epsilon=0.1$ we cannot determine $\tau$ since the correlation 
function with $t_w=2^{14}$ does not reach the value of $0.7$.  Anyhow 
here $\tau$ would be so huge to be completely incompatible with an 
$\epsilon^{-6}$ scaling.

When using a lower threshold (that should underestimate less the 
true correlation time) the situation is even more dramatic. In 
this case we can only determine $\tau$ for $\epsilon\ge 0.3$, and 
the deviation from an $\epsilon^{-6}$ scaling is more severe.

From this first, very naive analysis, we can already conclude that 
if we could define a correlation time $\tau$ for $\epsilon\to 0$ 
it would be growing far more dramatically than like 
$\epsilon^{-6}$.  The analysis of section (VI.A) \cite{CRISOB} is 
in this sense probably incorrect since the authors average the 
correlation function over small $t_w$ values (but they are working 
in conditions not identical to us, both because the exact form of 
the non hamiltonian contribution to the force and since they are 
at $T=0.5$).

Since the question of correlation times, of their scaling behavior 
and of the functional form of the time dependent correlation 
functions of the non ham\-il\-ton\-ian system is of crucial interest, we 
have decided to perform a careful analysis of this issue. The 
standard approach for trying to expose clearly an asymptotic 
exponential behavior is based on the definition of a time 
dependent effective correlation time $\tau_{eff}(t)$.
If we are in the asymptotic large time regime of a phase with an 
exponential decay of time-dependent correlation functions

\be
  c(t) \simeq A e^{-\frac{t}{\tau}}\ ,
\ee
and

\be
  \tau_{eff}(t) \equiv \Bigl(\frac{1}{t}
    \log\bigl (\frac{c(t)}{c(2t)}\bigr)\Bigr)^{-1} = \tau\ .
\ee
$\tau_{eff}(t) \to \tau$ at large times.

\begin{figure}
  \epsfxsize=400pt\epsffile{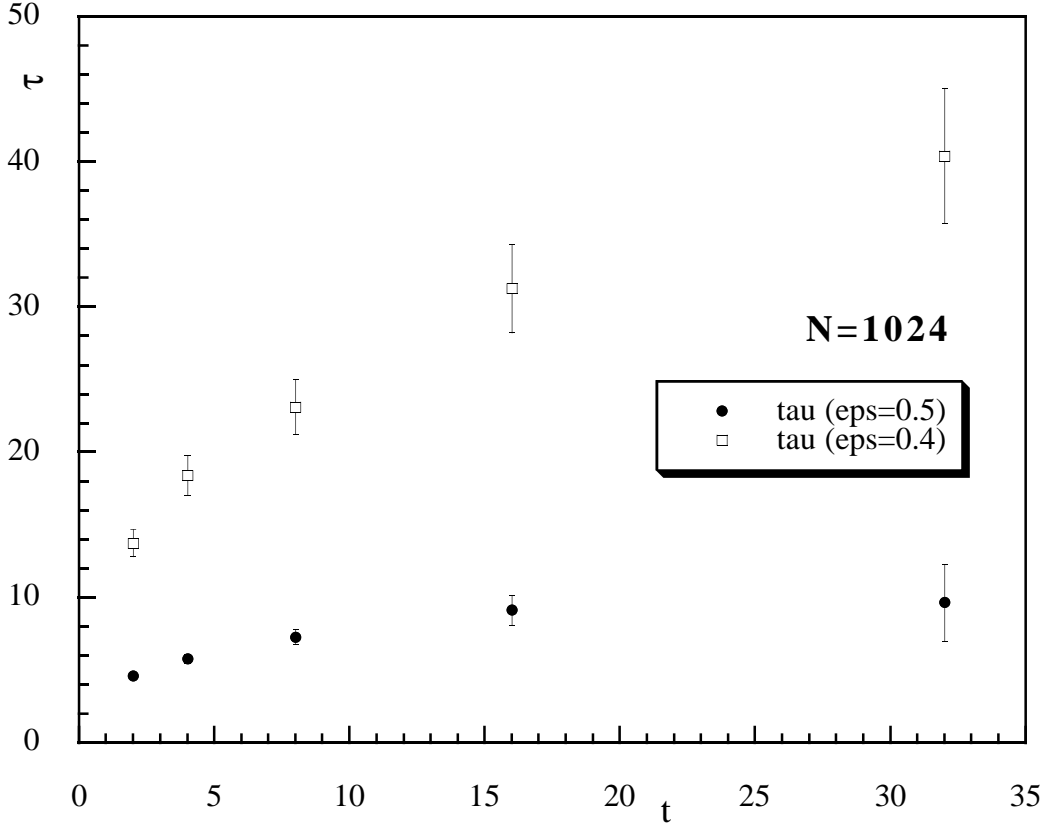}
  \caption[1]{
    $\tau_{eff}(t)$ versus $t$ for $\epsilon=0.5$ and $0.4$.
    }
  \protect\label{F-FIG12}
\end{figure}

We have hinted before that already figure \ref{F-FIG4} suggests 
that we cannot, from the data from our numerical simulations, 
exhibit a clean exponential behavior even at $\epsilon=.4$.  In 
fig.  \ref{F-FIG12} we plot $\tau_{eff}(t)$ versus $t$ for 
$\epsilon=0.5$ and $0.4$.  

We have averaged curves for different $t_w$ in the region where 
the dependence over $t_w$ is smaller than the statistical error 
($16$ for $\epsilon=0.5$, $64$ for $\epsilon=0.4$, $512$ for 
$\epsilon=0.3$, $204$8 for $\epsilon=0.2$, while for 
$\epsilon=0.1$, $0.0$ we have only selected the largest $t_w$).  We 
plot the larger $\epsilon$ values since for lower $\epsilon$ it is 
quite clear we are not observing at all an exponential decay.  
But, as we said, already for large $\epsilon$ the effective 
correlation time steadily increases, as a function of $t$, in the 
time region we can handle safely.  It is not clear from our data 
if $\tau_{eff}(t)$ is maybe reaching a plateau, but it is clear 
that where we can determine it with good precision it has not 
reached an asymptotic value.

We have also tried global fits to an exponential decay, by 
changing the number of data points used in the fit.  They are 
quite bad for all $\epsilon$ values (but maybe at $\epsilon=0.5$ 
where the noise threshold is reached after only a few data 
points). If discarding enough points close to the origin maybe an 
exponential fit is preferred at $\epsilon=0.5$ and $0.4$, while a 
power fit is preferred at $\epsilon=0.3$. For lower $\epsilon$ values 
one cannot find a simple behavior that fits well the data.

The evidence presented in this section does not completely clarify the 
main issue.  On the time scales we can observe there is still aging 
for small non hamiltonian perturbations, while for large perturbations 
aging disappears.  Still, even in the case of large perturbations a 
pure exponential behavior is slow to emerge.  Somehow it is clear that 
we are dealing with a system with a very complex dynamics, even if it 
is difficult to establish if we are dealing with a transient behavior 
or with an asymptotic effect.  But complexity is strong, and manifests 
itself with different signatures we have discussed in some details, in 
all the parameter range we have explored.  If we try to define a 
correlation time, even where it is not clear we could define one, it 
turns out to diverge faster than what a spherical spin model analogy 
would predict.  That could be connected to a signature of the replica 
symmetry breaking pattern of the Parisi solution.

Maybe the most important question we are not able to answer in a 
precise way is: is something drastic happening when going from large 
non hamiltonian perturbations (where we know that we will eventually 
get a non-aging behavior) to small perturbations?  It is very 
difficult to discriminate among a transient behavior on a very long 
time scale and a true asymptotic behavior.  If any, our feeling about 
this issue is that yes, things at $\epsilon=0.1$ are different from 
things at $\epsilon \ge 0.2$.  This is based both on the scaling of 
$\tau(\epsilon)$ we have discussed before (that is not compatible even 
with the very high power predicted by the spherical spin model 
solution), and by a hand-waving argument we give now.  Let us be very 
conservative and say that $\tau(\epsilon=0.2)$ is of order $300$ (this 
is by far a lower bound, and $\tau$ is probably, if any, far larger 
than that), and $\tau(\epsilon=0.1)>10000$ (this is obvious).  At 
$\epsilon=0.1$ for all times in our measurement windows ($O(2^{14})$) 
we see a very good aging.  Than we would expect that at $\epsilon=0.2$ 
at least for $t\ll 300$ we should have a transient aging, that could 
die out later on.  We do not have that at all.  Already at $t_w=2, 4$ 
aging curves at $\epsilon=0.2$ behave in a way that is dramatically 
different than for usual aging.  The argument of the transient 
behavior was used in \cite{CRISOA} to describe the situation at finite 
$\epsilon$, and it surely works for the spherical spin model: but here 
we have evidence that the argument does not apply.

Because of that, of the wrong scaling and of the further 
evidence we will present in the next section we cannot exclude that 
something changes at a critical value of $\epsilon$.

\section{Power Laws\protect\label{S-POWERL}}

Slow relaxation toward thermal equilibrium expectation values is one 
of the typical signatures of disordered systems. First Gardner, 
Derrida and Mottishaw \cite{GADEMO} have pointed out and quantified 
this kind of effect in spin glasses. Eisfeller and Opper 
\cite{EISOPP} have introduced a powerful dynamical functional method 
that allows to compute with good precision the power exponents at 
$T=0$. Ferraro \cite{FERRAR} has generalized this work to $T\ne 0$. 
Work based on numerical simulations \cite{ROSSET,BALDAS,BJMN} has 
made these computations detailed: one can determine with a reasonable 
accuracy power exponents even at $T\ne 0$, both for the remnant 
magnetization and for the energy decay.

\begin{figure}
  \epsfxsize=400pt\epsffile{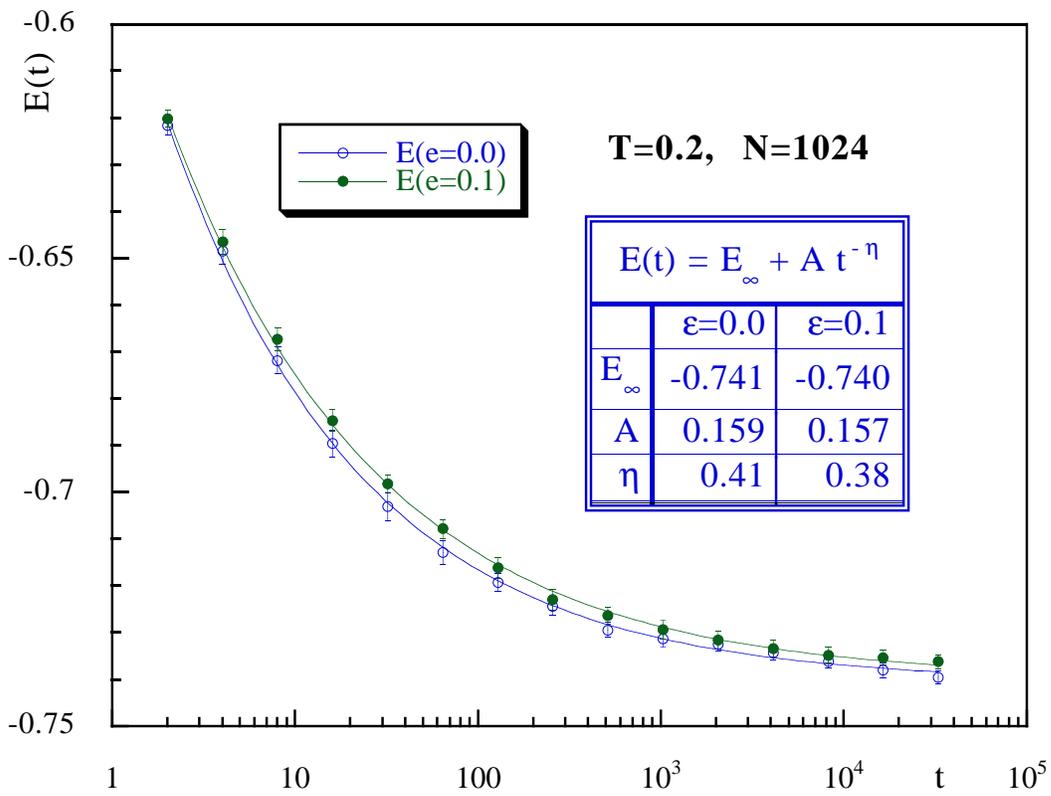}
  \caption[1]{
    $E(t)$ versus $t$ (linear-log scale) for $\epsilon=0.0$ and $0.1$.
    $N=1024$, $T=0.2$. The lines are for the best power fits.
    }
  \protect\label{F-FIGPL1}
\end{figure}

We have used the decay toward equilibrium of typical observable 
quantities (like the internal energy $E(t)$ and the squared overlap 
$q^2(t)$) as a probe of the existence of a complex behavior even for 
the non hamiltonian, $\epsilon\ne 0$ case. We have measured and tried 
to fit the time dependent internal energy

\be
  E(t) \simeq E_\infty + A t^{-\eta}\ ,
\ee
and analyzed the slow growth of $q^2(t)$ towards its equilibrium value.
On general ground we notice that the exponents we have determined for 
the energy are quite stable: they do not seem to depend much over the 
lattice size (we have checked different sizes) and over the time 
window we use to fit them.

First we present the results from our runs at $T=0.2$, the same we 
have discussed in the former section of this note. We start from 
$E(t)$. At $\epsilon=0.0$ and $\epsilon=0.1$ a power fit is perfect, 
while an exponential decay is clearly ruled out. In fig. \ref{F-FIGPL1} 
we plot the data with the best fits for $\epsilon=0.0$ and 
$\epsilon=0.1$. The error is here of $1$ to $3$ on the last digit. It 
is clear that the two sets of data are compatible in the statistical 
error: a small shift in the effective temperature of the system is the 
best way to explain the small shift in the data (the situation will 
be similar for higher $T$ values). Also at $\epsilon=0.2$ and $0.3$ we get a 
perfect power fit, with an exponent of $.39$ and $.48$ respectively, and 
an exponential behavior is ruled out. $\epsilon=0.4$ is an 
intermediate case, where both a power fit (that would give an exponent 
of $.63$) and an exponential fit are not very good. At $\epsilon=0.5$ 
the exponential fit is definitely better, and the energy clearly 
reaches its asymptotic value.

For zero and small $\epsilon$ we only see the very slow growth of 
$q^2(t)$, that is very far from its asymptotic value.  At 
$\epsilon=0.0$ and $\epsilon=0.1$ the slow growth is compatible with 
a logarithmic behavior.  For higher $\epsilon$ values one starts to 
see $q^2(t)$ approaching an asymptotic value, but in this case, even 
when the asymptotic value is very clear, a power fit does not work 
well.  It is interesting to notice that somehow the functional form of 
the time dependence of $q^2(t)$ is very different from the one of the 
energy $E(t)$, where the power law is very clear.  As we will also 
show for $T=0.5$ with fig.  (\ref{F-FIGPL3}) the best way to describe 
the behavior of $q^2(t)$ would be by a very slow logarithmic growth 
that stops abruptly after reaching its finite volume asymptotic 
value.

\begin{figure}
  \epsfxsize=400pt\epsffile{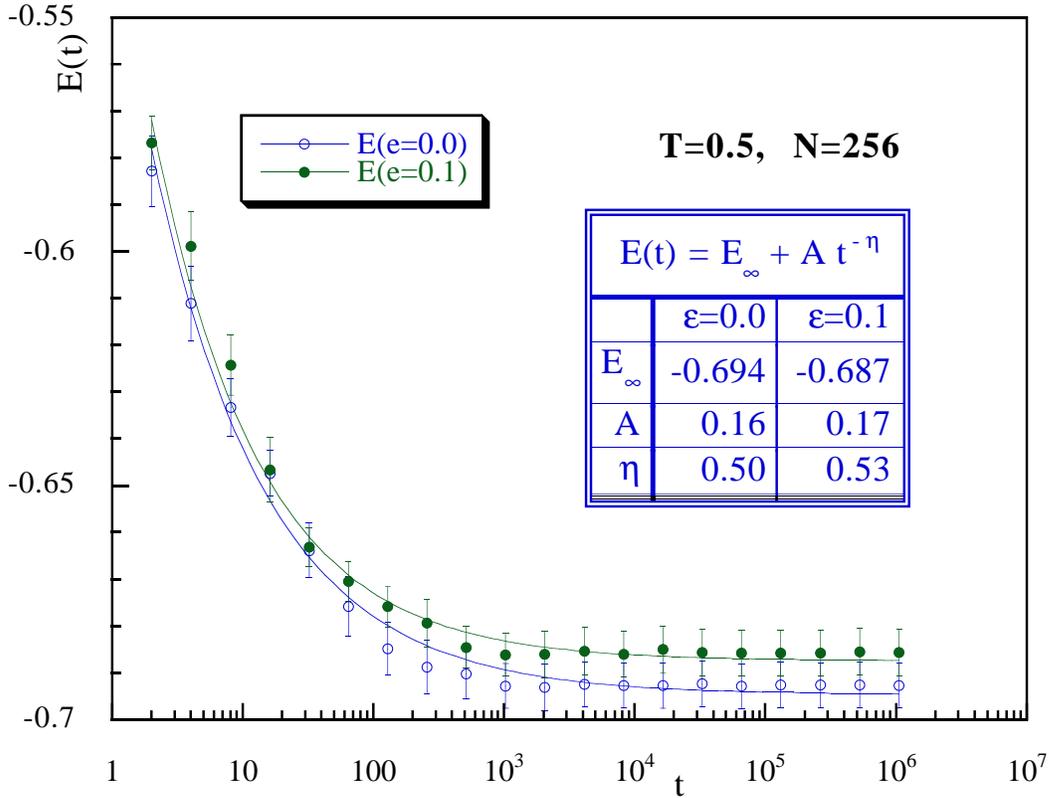}
  \caption[1]{
    $E(t)$ versus $t$ (linear-log scale) for $\epsilon=0.0$ and $0.1$.
    $N=256$, $T=0.5$. The lines are for the best power fits.
    }
  \protect\label{F-FIGPL2}
\end{figure}

We have also studied the system at higher $T=0.5$ and smaller volume 
$N=256$, by running longer simulations ($10^6$ steps, and $10$ samples 
for each $\epsilon$ value).  That has been done in order to reach 
equilibrium at $\epsilon=0.0$ and compare with finite $\epsilon$ (to 
check, for example, if the finite $\epsilon$ system converges to an 
effective stable Boltzmann-like probability distribution, see 
(\ref{S-EQUILI}).  The situation is very similar to the case of 
$T=0.2$, with the difference that here the energy and the $q^2$ 
plateau's are very clear already at $\epsilon=0.0$ (equilibrium for 
both energy and $q^2$ are apparently reached after $O(10^3)$ steps), 
and the exponents of the power decay are higher (since we are at 
higher $T$).  We show in fig.  (\ref{F-FIGPL2}) $E(t)$ at 
$\epsilon=0.0$ and $0.1$ with the best power fit (errors are up to 
$O(5)$ on the last digit).  Again the results are very similar, and the 
power fit is very good.  An exponential fit is not able to describe 
the data.  Here already at $\epsilon=.2$ convergence to equilibrium is 
very fast and an exponential fit works reasonably.  At $\epsilon=0.4$ 
the best exponent of the power fit is close to one.  At $\epsilon=0.5$ 
the exponential fit is perfect.

\begin{figure}
  \epsfxsize=400pt\epsffile{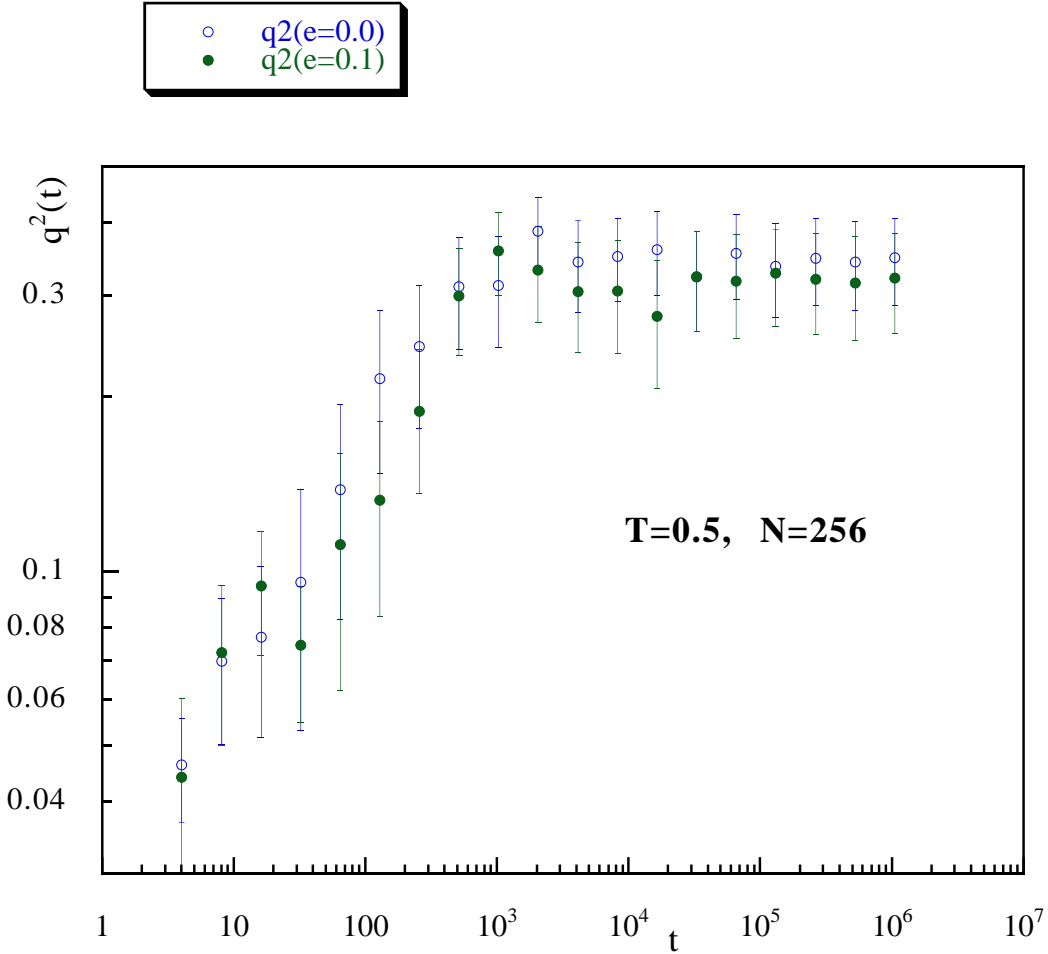}
  \caption[1]{
    $q^2(t)$ versus $t$ (linear-log scale) for $\epsilon=0.0$ and $0.1$.
    $N=256$, $T=0.5$.
    }
  \protect\label{F-FIGPL3}
\end{figure}

$q^2(t)$ converges to a clear plateau, but as we said it is difficult 
to find a correct functional form to describe such time dependence.  
In fig.  (\ref{F-FIGPL3}) we show $q^2(t)$ at $\epsilon=0.0$ and 
$0.1$, where a power fit does not work well. The situation is 
qualitatively very similar at higher $\epsilon$ values, but $q^2(t)$ 
reaches lower plateau values.
At $\epsilon=0.4$ the asymptotic value of $q^2$ is very small.  At 
$\epsilon=0.5$ the asymptotic value of $q^2$ is $0.02$, i.e.  fully 
compatible with zero.

To sum it up, slow convergence to equilibrium is clear for different 
values of $T<T_c^{(\epsilon=0.0)}$. We can never see, in the time 
ranges we investigate, any significant difference when going from 
$\epsilon=0.0$ to $\epsilon=0.1$. It also has to be noticed that in 
the case where we reach thermal equilibrium for $\epsilon$ not too 
large the expectation value of $q^2$ at equilibrium is far larger 
than a pure finite size contamination: for $N=256$ $\langle q^2 
\rangle$ is clearly non zero in a large $\epsilon$ range. We will try 
to understand if this is or not a finite size effect in the next section.

\section{Equilibrium\protect\label{S-EQUILI}}

In the former two sections we have analyzed the dynamical properties 
of the non hamiltonian dynamics. In this section we have taken the 
complementary point of view and, by studying small lattices on longer 
time scales we have investigated the equilibrium properties of the 
model. Even if the model is not defined from an Hamiltonian we can 
define its equilibrium properties from the large time limit of the 
dynamics. We have measured the probability distribution of the 
overlap $P(q)$, and compared the SK model to the non hamiltonian 
dynamics. We have systematically analyzed the dependence of the 
equilibrium properties from the finite volume (up to the largest volume 
on which we have been able to thermalize the system, see later).

We have worked at $T=.5$, running $10^6$ full sweeps of the lattice 
and using the second two thirds of the sweeps for measuring 
equilibrium properties. We have selected, as before, $\epsilon$ going 
from $0$ to $0.5$ with increments of $0.1$.

We have used $N=64$, $128$ and $256$ (respectively with $20$, $10$ and 
$10$ samples for each $\epsilon$ value).  In all case we have checked 
that all the relevant observable quantities (for example $E(t)$, 
$q^2(t)$) have reached a very clear plateau, where they are stable in 
all of the measurement region.  We have also analyzed directly the 
sample dependent probability distributions $P_J(q)$ checking the 
symmetry sample by sample (a very strong check).  At $\epsilon=0$, 
where the thermalization is more difficult, the $P_J(q)$ is very 
symmetric at $V=64$ on all samples.  At $V=128$ there is again a very 
good symmetry (the maximum discrepancy is of order of $25$ per cent of 
the double peak height.  At $V=256$ some of the samples have a quite 
asymmetric $P(q)$, but for all $10$ systems it looks very plausible 
that true equilibrium has been reached: the average $P(q)$ is nicely 
symmetric. We would not have succeeded to thermalize larger systems. 
  
\begin{figure}
  \epsfxsize=388pt\epsffile{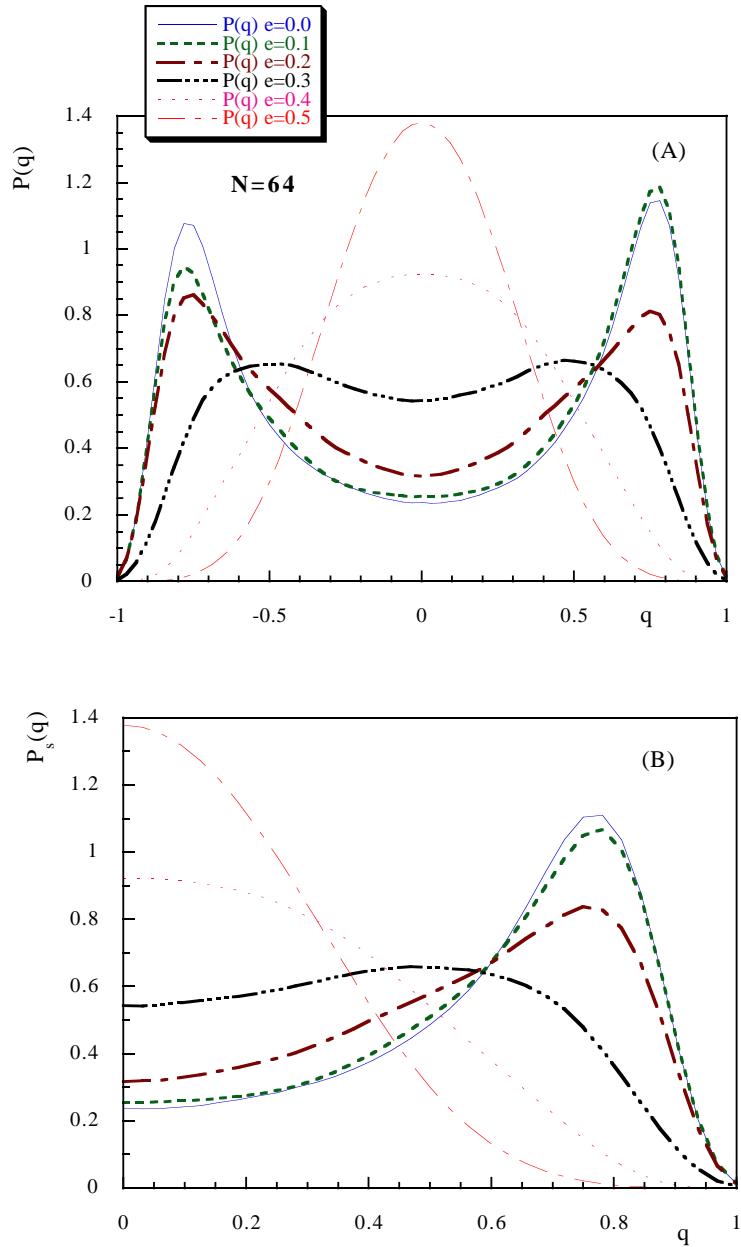}
  \caption[1]{$P(q)$ versus $q$ for different $\epsilon$ values, $N=64$. 
  In (A) the full $P(q)$, where the quality of the symmetry under $q 
  \to -q$ gives a measure of how good our thermalization was. In (B) 
  the symmetrized $P_s(q)$.
  }
  \protect\label{F-FIGEQ1}
\end{figure}
  
\begin{figure}
  \epsfxsize=388pt\epsffile{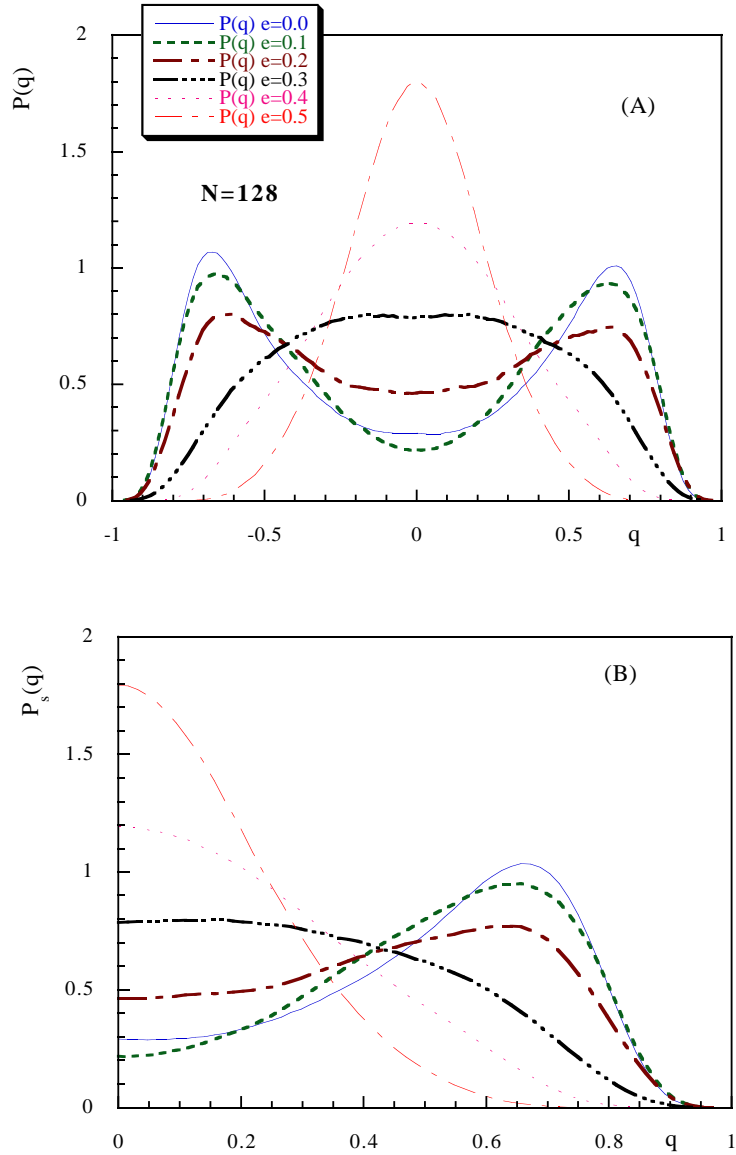}
  \caption[1]{
  As in figure (\protect\ref{F-FIGEQ1}), but $N=128$.
  }
  \protect\label{F-FIGEQ2}
\end{figure}
  
\begin{figure}
  \epsfxsize=388pt\epsffile{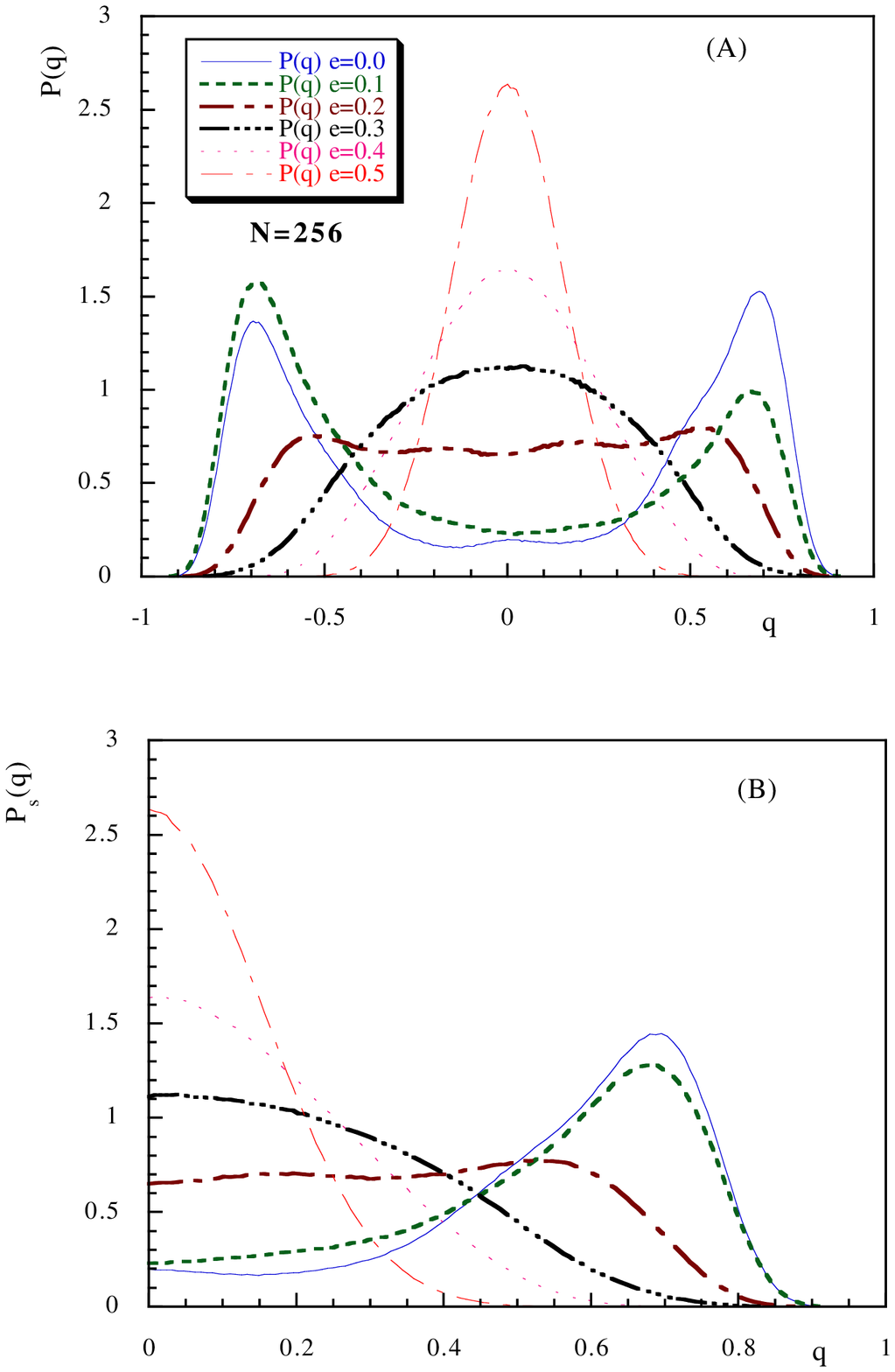}
  \caption[1]{
  As in figure (\protect\ref{F-FIGEQ1}), but $N=256$.
  }
  \protect\label{F-FIGEQ3}
\end{figure}
  
In figures (\ref{F-FIGEQ1})-(\ref{F-FIGEQ2}) we plot $P(q)$ for 
different $\epsilon$ and $N$ values (always at $T=0.5$). It is clear 
the double peak structure of the Parisi broken phase of the SK model, 
and that for high $\epsilon$ values one gets a trivial distribution 
centered around $q=0$.

It is more interesting to follow for example the $\epsilon=0.3$ case as 
a function of $N$. Here from a (already quite soft) double peak 
structure at $N=64$ one goes to a broad peak around $q=0$ at $N=256$.
At $\epsilon=0.2$ there is the same kind of effect: a strong double 
peak at $N=64$ softens at $N=128$. At $N=256$ we are left with a flat 
plateau including $q$ values going from $-.5$ to $.5$.

In the case $\epsilon=0.1$ once again we get results that are very 
similar to the ones we get for the pure SK model. In our $N$ range we 
cannot observe any systematic effect.

\begin{figure}
  \epsfxsize=400pt\epsffile{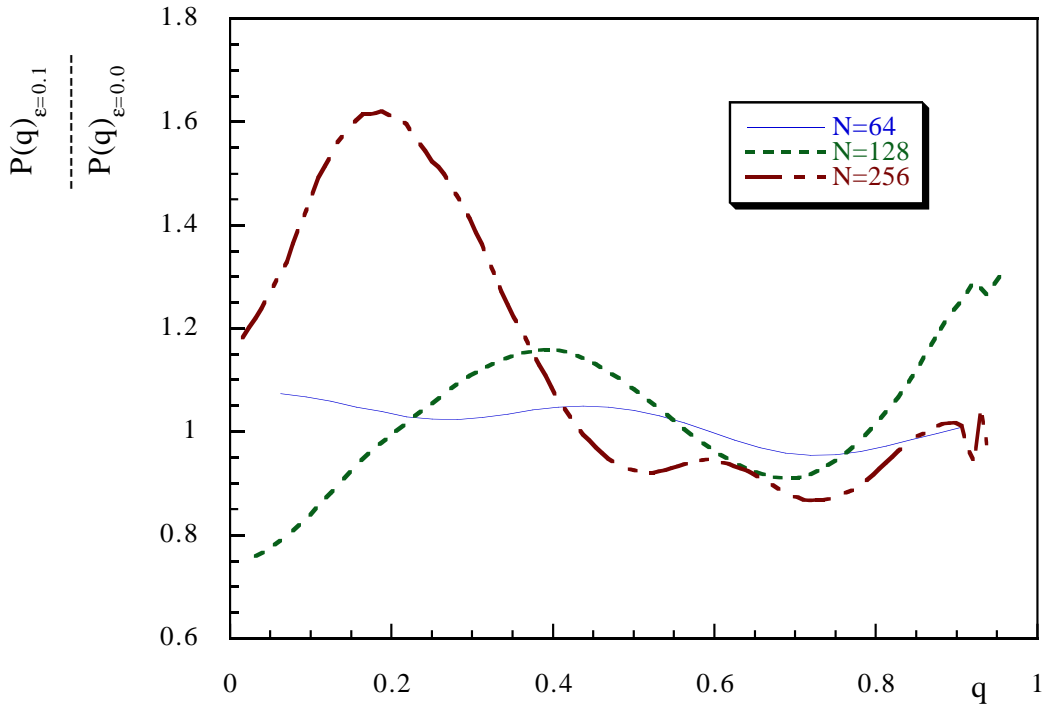}
  \caption[1]{
  The ratio of $P(q)$ at $\epsilon=0.0$ and $\epsilon=0.1$ as a 
  function of $q$, for different $N$ values.
  }
  \protect\label{F-FIGEQ4}
\end{figure}

For making this point more clear we plot in fig. (\ref{F-FIGEQ4}) the 
ratio of the $P(q)$ for the $\epsilon=0.1$ model and the pure SK 
model as a function of $q$, for the three $N$ values we have analyzed.
A part from large fluctuations we cannot see any systematic trend. For 
small $q$ on the large lattice we have a quite large ratio, but the 
statistical undetermination is in this case large.

In this section we have been exploring features of the model that are 
different from the dynamical issues we were discussing before. Here we 
are discussing about equilibrium properties: the system could very 
well have a complex dynamics on divergent time scales but a trivial 
large time limit. Still, what we find is that again, on the volume
scales we are able to disentangle numerically, there is no difference 
among the SK model and the one with a small non hamiltonian 
perturbation. Also we notice that at least for large perturbations the 
small lattices fake a non-trivial structure, that disappears in the 
infinite volume limit. The same effect could make trivial the theory 
with small perturbations on very large volume, but here we cannot see 
such an effect.

\section{Conclusions\protect\label{S-CONCLU}}

Our numerical simulations surely show that the non hamiltonian systems 
we have studied have a very interesting, complex behavior, and show 
that the spherical spin model analogy is probably not all of the 
story.

We have studied aging. We have found that on the time scales we could 
investigate (that are, one should not forget, of the order of the 
largest time scales that have been used to claim numerically that the 
pure model undergoes aging) systems with small perturbations age, 
while systems with a large non hamiltonian term have a conventional 
behavior (with some care to be used when discussing the intermediate 
$\epsilon$ case, $\epsilon=0.2$ for us). Our data do not look 
compatible with the $\epsilon^{-6}$ scaling one finds for the 
spherical spin model \cite{CRISOA,CRISOB}. Also the aging systems look 
different from the non-aging ones on all time scales, making the 
possibility of a transient behavior less favored.

We have studied the time dependence of observables like the internal 
energy $E(t)$ or $q^2(t)$. We have found clear power law decays for 
small perturbations.

At last we have studied equilibrium. Here we are looking at a regime 
that is very different from the one discussed before. Also here we 
have seen that for small perturbations one finds results that are very 
similar to the ones of the pure model, but we have also seen that for 
larger $\epsilon$ small lattices do produce fake double peak 
structures in the probability distribution of the overlap.

There is space, as many times is the case, for more analytical and 
numerical work.

\section*{Acknowledgments}

We warmly thank David Dean and Daniel Stariolo for many useful 
exchanges about the subject.  We are indebted to Leticia Cugliandolo 
and Jorge Kurchan for informing us about the results of ref.  
\cite{CKLP} prior to publication, and for a very explicative and 
helpful conversation.  We acknowledge useful 
conversations with Giorgio Parisi, Paola Ranieri and Juan Ruiz-Lorenzo.

\end{document}